\documentclass{elsart}

\usepackage{graphics}
\usepackage{epsfig}

\usepackage{amssymb}

\begin{document}

\begin{frontmatter}



\title{Self-Organized Criticality and Stock Market Dynamics: an Empirical Study.}

\author{M. Bartolozzi$^{a}$, D. B. Leinweber$^{a}$,
A. W. Thomas$^{a,b}$}

\address{$^a$Special Research Centre for the Subatomic Structure of Matter (CSSM) 
and Department of Physics,
 University of Adelaide, Adelaide, SA 5005, Australia \\ 
$^b$Jefferson Laboratory, 12000 Jefferson Ave., Newport News, VA 23606, USA}

\begin{abstract}
The Stock Market is a complex self-interacting system, characterized by 
intermittent behaviour. Periods of high activity alternate with
periods of relative calm. In the present work we investigate empirically
the possibility that the market is in a self-organized critical
state (SOC). A wavelet transform method is used in order to separate high 
activity periods, related to the avalanches found in sandpile models, from quiescent.
 A statistical analysis of the 
filtered data shows a power law behaviour in the avalanche size, duration 
and laminar times. The memory process, implied by the power
law distribution of the laminar times, is not consistent with 
classical conservative models for self-organized criticality. We
argue that a ``near-SOC'' state or a time dependence in the driver,
which may be chaotic, can explain this behaviour. 
\end{abstract}

\begin{keyword}
Complex Systems \sep Econophysics \sep Self-Organized Criticality \sep Wavelets 
\PACS 
\end{keyword}
\end{frontmatter}

\section{Introduction}
 
Since the publication of the articles of Bak, Tang and 
Wiesenfeld (BTW)~\cite{Bak8788},
the concept of self-organized criticality (SOC) has been invoked  
to explain the dynamical behaviour of many complex systems, from physics
to biology and the social sciences~\cite{Jensen,Turcotte99}.
The key concept of SOC is that complex systems, that is systems constituted
by many interacting elements, although obeying different
microscopic physics, may exhibit similar dynamical
behaviour. In particular, the statistical properties of these systems 
can be described by power laws, reflecting a lack of
any characteristic scale. These features are equivalent 
to those of physical systems
during a phase transition, that is at the critical point.
It is worth emphasizing that the original idea~\cite{Bak8788}
was that the critical state was reached ``naturally'' , without 
any external tuning. This is the origin of the adjective {\em self-}organized.
 In reality a certain degree of tuning is
necessary: implicit tunings like local conservation laws 
and specific boundary
conditions seem to be important ingredients for the appearance of 
power laws~\cite{Jensen}.

The classical example of a system exhibiting SOC behaviour 
is the 2D sandpile model~\cite{Bak8788,Jensen,Turcotte99}. Here the cells
of a grid are randomly filled, by an external random driver, with ``sand''. 
When the gradient between two 
adjacent cells exceeds a certain threshold a redistribution of the
sand occurs, leading to more instabilities and further redistributions.
The benchmark of this system, indeed of all systems exhibiting SOC, is 
that the distribution of the avalanche sizes, their duration and
the energy released, obey power laws.
 
The framework of self-organized criticality 
has been claimed to play an important role in solar flaring~\cite{flares},
space plasmas~\cite{spplasma} and earthquakes~\cite{earthquake} 
in the context of both astrophysics and geophysics. 
In the biological sciences, SOC, has been related, for example, with 
biodiversity and evolution/extinction~\cite{Bak93}.
Some work has also been carried out in the social sciences. 
In particular, traffic flow and traffic jams~\cite{traffic},
 wars~\cite{Roberts98} and  
stock-market~\cite{Turcotte99,Bak93b,Bak97,Feigenbaum03} 
dynamics have been studied.
A more detailed list of subjects and references related to SOC can 
be found in the review paper of Turcotte~\cite{Turcotte99}.

\begin{figure}
\centerline{\epsfig{figure=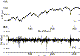,height=8cm, width=10cm}}
\caption{ Sample of the tick-by-tick time series of the Nasdaq100(a),
 as well as the corresponding returns (b).}
\label{fig1}
\end{figure}

In the present work we will provide empirical evidence for connections
between self-organized criticality and the stock market, considered as
a complex system constituted of many interacting individuals.
We analyze the tick-by-tick behaviour of the
Nasdaq100 index, $P(t)$, from 21/6/1999 to 19/6/2002 for a total of $2^{19}$
data. A sample of this data is illustrated in Fig.~\ref{fig1}(a).
 In particular, we study the logarithmic returns of
this index, which are defined as
$R(t)=\ln(P(t+1))-\ln(P(t))$ and plotted in Fig.~\ref{fig1}(b).

To examine the extent to which our findings apply 
to other stock market indices we also studied the S\&P ASX50 (for
the Australian stock market) at intervals of 30 minutes over
the period 20/1/1998 to 1/5/2002, for a total of  $2^{14}$ data
points.
Possible differences between daily and high frequency data have 
also been taken into consideration though the analysis of the Dow Jones daily 
closures from 2/2/1939 to 13/4/2004. The results are presented 
in Sec.~\ref{sec:dataAnalysis}.

From a visual 
analysis of the time series of returns, Fig.~\ref{fig1}(b), we observe
long periods of relative tranquility, characterized by small fluctuations,
and periods in which the index goes through very large fluctuations, 
equivalent to avalanches, clustered in relatively short time intervals.
These may be viewed as a consequence of a build-up process leading
the system to an extremely unstable state. Once this critical point
has been reached, any small fluctuation can, 
in principle, trigger a chain reaction, 
similar to an avalanche, which is needed to stabilize the system again.

\section{Wavelet Method} 
\label{sec:wavelet}

\begin{figure}
\vspace{1cm}
\centerline{\epsfig{figure=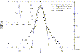,height=8cm, width=10cm}}
\caption{ PDF of the logarithmic returns of the Nasdaq100 index before (triangles)
and after filtering (circles), with $C=2$. The original time series is reduced 
to the level of noise. A Gaussian distribution 
is plotted for comparison. The insert shows 
the fourth member of the Daubechies wavelets used in the filtering.}
\label{fig2}
\end{figure}

With the recent development of the interdisciplinary area of
complexity, many physicists have started  to study the dynamical
properties of stock markets~\cite{mantegna,paul}.  Empirical results
have shown that the time series of financial  returns show a behaviour
similar to  hydrodynamic turbulence~\cite{Ghashghaie96,Mantegna97} --
although differences have also been pointed
out~\cite{Mantegna97}. Both the spatial velocity fluctuations in
turbulent flows and the stock market returns show an intermittent
behaviour, characterized by broad tails in the probability
distribution function (PDF), and a non-linear multifractal
spectrum~\cite{Ghashghaie96}.  The PDF for the normalized logarithmic
returns,
\begin{equation}
r(t)=\frac{R(t)-\langle R(t) \rangle_{l}}{\sigma(R(t))},
\end{equation}
where $\langle\ldots\rangle_{l}$ is the average over the length of the sample,
$l$, and $\sigma$ the standard deviation, is plotted in Fig.~\ref{fig2}. 
The departure from a Gaussian behaviour is evident, in particular,
in the peak of the distribution and in the broad tails, 
which are related to extreme events. 

The empirical analogies between turbulence and the stock market may 
suggest the existence of a temporal information 
cascade for the latter~\cite{Ghashghaie96}.  This
is equivalent to assuming that various traders require different
information according to their specific strategies.
In this way different time scales become involved in the trading
process.
In the present work we use a wavelet method in order to study multi-scale
market dynamics.

The wavelet transform is a
relatively new tool for the study of intermittent and multifractal 
signals~\cite{Farge92}. The approach
enables one to decompose the signal in terms of scale and time units
and so to separate its coherent parts -- that is, the bursty periods related 
to the tails of the PDF -- from the noise-like background, thus enabling an 
independent study of the intermittent and the 
quiescent intervals~\cite{Farge99}. 

The continuous wavelet transform (CWT) is defined 
as the scalar product of the analyzed 
signal, $f(t)$, at scale $\lambda$ and time $t$, with a real or complex 
``mother wavelet'', $\psi(t)$:
\begin{equation}
W_{T}f(t)=\langle f,\psi_{\lambda,t}\rangle=\int f(u) \bar{\psi}_{\lambda,t}(u)du=
\frac{1}{\sqrt{\lambda}}\int f(u) \bar{\psi}(\frac{u-t}{\lambda})du.
\label{cwt}   
\end{equation}   
The idea behind the wavelet transform is similar to that of   
windowed Fourier analysis and it can be shown that the
scale parameter is indeed inversely proportional to
the classic Fourier frequency. The main 
difference between the two techniques lies in the resolution in the
time-frequency domain. In the Fourier analysis the 
resolution is scale independent, leading to aliasing of high and 
low frequency components that do not fall into 
the frequency range of the window. However
in the wavelet decomposition the
resolution changes according to the scale (i.e. frequency).
At smaller scales the temporal resolution increases at the expense of frequency
localization, while for large scales we have the opposite. For this
reason the wavelet transform is considered a sort of mathematical 
``microscope''. While the Fourier analysis is still an appropriate method
for the study of harmonic signals, where the information is equally
distributed, the wavelet approach becomes fundamental when the
signal is intermittent and the information localized. 

The CWT of Eq.(\ref{cwt}) is a powerful tool to graphically identify
coherent events, but it contains a lot of redundancy in the
coefficients. For a time series analysis it is often preferable to 
use a discrete wavelet transform (DWT). The DWT can be seen as a 
appropriate sub-sampling of Eq.(\ref{cwt}) using dyadic scales.  
That is, one chooses $\lambda=2^{j}$, for $j=0,...,L-1$, 
where $L$ is the number of
scales involved, and the temporal coefficients 
are separated by multiples of $\lambda$ for each dyadic scale, $t=n 2^{j}$,
with $n$ being the index of the coefficient at the $j$th scale. 
The DWT coefficients, $W_{j,n}$, can then be expressed as
\begin{equation}
W_{j,n}=\langle f,\psi_{j,n}\rangle=2^{-j/2}\int f(u) \psi(2^{-j}u-n) du,
\label{dwt}   
\end{equation}     
where $\psi_{j,n}$ is the discretely scaled and shifted version of the 
mother wavelet. The wavelet coefficients are a measure of the
correlation between the original signal, $f(t)$, and the mother
wavelet, $\psi(t)$  at scale $j$ and time $n$.
In order to be a wavelet, the function  $\psi(t)$ must satisfy some conditions.
First it has to be well localized in both real and Fourier
space and second the following relation
\begin{equation}
C_{\psi}=2\pi \int_{-\infty}^{+\infty}\frac{|\hat{\psi}(k)|^{2}}{k}dk<\infty,
\label{admiss}
\end{equation}
must hold, where $\hat{\psi}(k)$ is the Fourier transform of $\psi(t)$. 
The requirement expressed by Eq.(\ref{admiss}) is called {\em admissibility}
and it guarantees the existence of the inverse wavelet transform. 
The previous conditions are generally satisfied if the mother wavelet
is an oscillatory function around zero, with a rapidly decaying envelope.
Moreover, for the DWT, if the set of the mother wavelet and 
its translated and scaled copies form
an orthonormal basis for all functions having a finite squared modulus, 
then the energy of the starting signal is
conserved in the wavelet coefficients. This property 
is, of course, extremely important
when analyzing physical time series~\cite{Kovacs01}.
More comprehensive discussions on the wavelet
properties and applications are
given in  Refs.~\cite{Daubechies88} and \cite{Farge92}.
Among the many orthonormal bases known,
in our analysis we use the fourth member of the 
Daubechies wavelets~\cite{Daubechies88}, shown in the insert of Fig.~\ref{fig2}.
The spiky form of this wavelet insures a strong correlation 
for the bursty events in the time series. The following method of 
analysis has also been tested with other wavelets and the results are qualitatively 
unchanged.

\begin{figure}
\centerline{\epsfig{figure=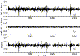,height=8cm, width=10cm}}
\caption{ A sample of the original time series
 of logarithmic returns for the Nasdaq100 is shown in (a), 
same as Fig.~\ref{fig1}(b). The filtered version is shown in (b). The noise-like
behaviour of this time series is evident. The residual time series is shown 
in (c). This corresponds to the high activity periods of the time series, 
related to the broad wings of the PDF. The cut-off parameter in this case 
is $C=2$.}
\label{fig3}
\end{figure}

The importance of the wavelet transform in the study of turbulent
signals lies in the fact that the
large amplitude wavelet coefficients 
are related to the extreme events in the tails of the PDF, while
the laminar or quiescent periods are related to the 
ones with smaller amplitude~\cite{Kovacs01}. 
In this way it is possible to define a criterion whereby one can filter the
time series of the coefficients depending on the specific needs.
In our case we adopt the method used in Ref.~\cite{Kovacs01} and
originally proposed by Katul et al~\cite{Katul94}.
In this method wavelet coefficients
that exceed a fixed threshold are set to zero, according to
\begin{equation}
\tilde{W}_{j,n}=\left \{  \begin{array}{ccc} W_{j,n} & {\rm if} &
 W^{2}_{j,n}<C\cdot\langle W^{2}_{j,n}\rangle_{n},\\
0 & {\rm otherwise},
\end{array} \right.
\end{equation}
here $\langle \ldots \rangle_{n}$ denotes the average over the time
parameters at a certain scale and $C$ is the threshold coefficient.
In the next section we will see that the precise value of the 
parameter $C$ is not critical for our analysis. However it is
possible to tune $C$ such that only Gaussian noise is filtered.
Once we have filtered the wavelet coefficients $\tilde{W}_{j,n}$ we perform
an inverse wavelet transform, obtaining a {\em smoothed} 
version, Fig.~\ref{fig3}(b), of the original time series, Fig.~\ref{fig3}(a). 
The residuals of the original time 
series with the filtered one correspond to the bursty periods 
which we aim to study, Fig.~\ref{fig3}(c). 
%

\section{Data Analysis}
\label{sec:dataAnalysis}

In the previous section we have introduced the wavelet method in
order to distinguish periods of high activity and periods of low or
noise-like activity. The results are shown in Fig.~\ref{fig3} for $C=2$. 
In order to choose an appropriate cut-off for the wavelet energy,
that is to fix a proper $C$, we tune this parameter until the statistics
on the kurtosis and the skewness of the filtered time series approach the
noise levels, namely 3 and 0 respectively.
Once we have isolated the noise part of the time series we 
are able to perform a reliable statistical analysis on the {\em coherent events}
of the residual time series, Fig.~\ref{fig3}(c). In particular, we define 
coherent events as the periods of the residual
time series in which the volatility, $v(t)\equiv |R(t)|$, is above a
small threshold, $\epsilon \approx 0$. 
The smoothing procedure is emphasized by the change in the PDFs before 
and after the filtering -- as shown in Fig.~\ref{fig2}. From 
this plot it is clear how the broad tails, 
related to the high energy events that we
want to study, and the associated central peak
 are cut-off by the filtering procedure. The
filtered time series is basically a Gaussian, related to a noise process.

A parallel between avalanches in the classical sandpile models 
(BTW models) exhibiting SOC~\cite{Bak8788} and the previously defined
coherent events in the stock market is straightforward.
In order to test the relation between the two, we make use of some
properties of the BTW models. In particular, we use the fact that 
the avalanche size distribution and the avalanche duration 
are distributed according to power laws, while the laminar, or waiting  
times between avalanches are exponentially distributed, reflecting
the lack of any temporal correlation between them~\cite{Boffetta99,Wheatland98}.
This is equivalent to stating that the triggering process has no memory.

\begin{figure}
\centerline{\epsfig{figure=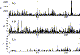,height=8cm, width=10cm}}
\caption{ Time series for the avalanche volumes, $V$, for the Nasdaq100, (a);
 duration, $D_t$, of the avalanches, (b); and laminar times, $L_t$ (c). 
The plots are obtained using $C=2$ as the filtering parameter.}
\label{fig4}
\end{figure}

Similarly to the dissipated energy in a turbulent flow, we define
an avalanche, $V$, in the market context as the integrated
value of the squared volatility, over each coherent event of 
the residual time series.
The duration, $D_{t}$, is defined as the interval of time 
between the beginning and
the end of a coherent event, while the laminar time, $L_{t}$, is the
time elapsing between the end of an event and the beginning of the next one.
The time series for $V$, $D_{t}$ and $L_{t}$ are plotted 
in Fig.~\ref{fig4} for $C=2$.
%

\begin{figure}
\vspace{1cm}
\centerline{\epsfig{figure=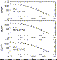,height=8cm, width=10cm}}
\caption{ Probability distribution function for the avalanche 
sizes tested against
several values of $C$. The power law behaviour 
is robust with respect to this parameter.}
\label{fig5}
\end{figure}

\begin{figure}
\vspace{2cm}
\centerline{\epsfig{figure=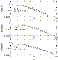,height=8cm, width=10cm}}
\caption{ The duration of the high activity periods show a power law distribution,
independent of the value of $C$.}
\label{fig6}
\end{figure}

\begin{figure}
\vspace{1cm}
\centerline{\epsfig{figure=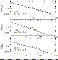,height=8cm, width=10cm}}
\caption{ Power law distribution of laminar times for different
values of $C$.}
\label{fig7}
\end{figure}

The results for the statistical analysis for the Nasdaq100 index 
are shown in Figs.~\ref{fig5}, 
\ref{fig6} and \ref{fig7}, respectively, for the avalanche size,
duration and laminar times. The robustness of our method has been tested
against the energy threshold, we perform the same analysis 
with different values of $C$.

A  power law relation is clearly evident for  
all the quantities investigated, largely independent of the
specific value of $C$. At this point is important to stress the 
difference in the distribution of laminar times between the BTW 
model and the data analyzed. As explained previously, the BTW model shows  
an exponential distribution for the latter,
derived from a Poisson process with no memory~\cite{Boffetta99,Wheatland98}. 
The power law distribution found for the stock market instead implies
the existence of temporal correlations between coherent events.
This empirical result rules-out
the hypothesis that the stock market is in a 
SOC state, at least in relation to the classical sandpile models.
 
\begin{figure}
\vspace{2cm}
\centerline{\epsfig{figure=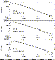,height=8cm, width=10cm}}
\caption{ Probability distribution function for the avalanche 
sizes for the S\&P ASX50, from 20/1/1988 to 1/5/2002.}
\label{fig8}
\end{figure}

\begin{figure}
\vspace{2cm}
\centerline{\epsfig{figure=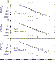,height=8cm, width=10cm}}
\caption{ Distribution of the duration of the {\em coherent events} 
for the S\&P ASX50.}
\label{fig9}
\end{figure}

\begin{figure}
\vspace{1cm}
\centerline{\epsfig{figure=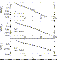,height=8cm, width=10cm}}
\caption{ Distribution of laminar times for S\&P ASX50 index.}
\label{fig10}
\end{figure}

In order to extend the study of the avalanche behaviour to different markets,
 we perform the same analysis over the 30 minute 
returns for the S\&P ASX50. The results are shown in Figs.~\ref{fig8},~\ref{fig9}
 and ~\ref{fig10}. While the power law scaling for the
laminar times is still very clear, the power law for the other quantities 
is to less precise, perhaps reflecting a different underlying 
dynamics compared to the Nasdaq100. On the other hand it is also important to stress 
the difference in length of the two time
series analyzed. While for the Nasdaq100 we used $2^{19}$ data
points, only $2^{14}$ were available for the S\&P ASX50, making the 
first study statistically more reliable.  

\begin{figure}
\vspace{1cm}
\centerline{\epsfig{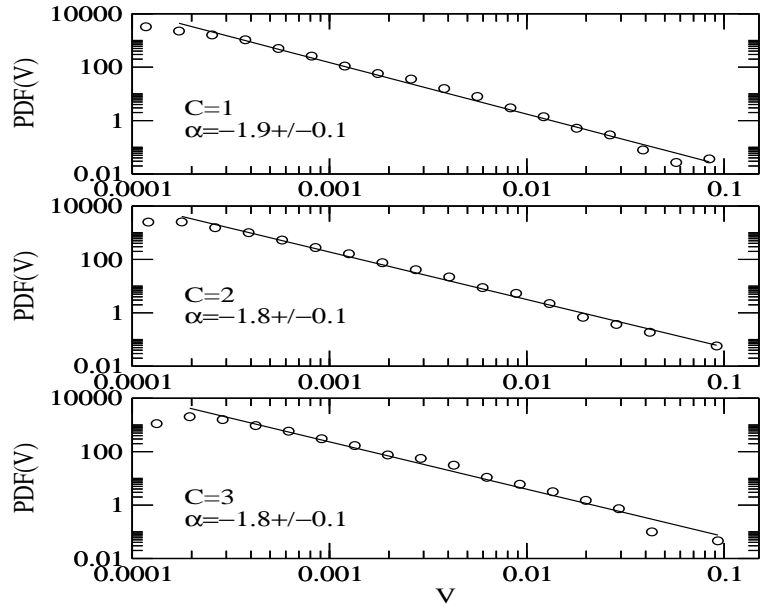}}
\caption{ Probability distribution function for the avalanche 
sizes for the Dow Jones daily closures, from 2/2/1939 to 13/4/2004.}
\label{fig11}
\end{figure}

\begin{figure}
\vspace{2cm}
\centerline{\epsfig{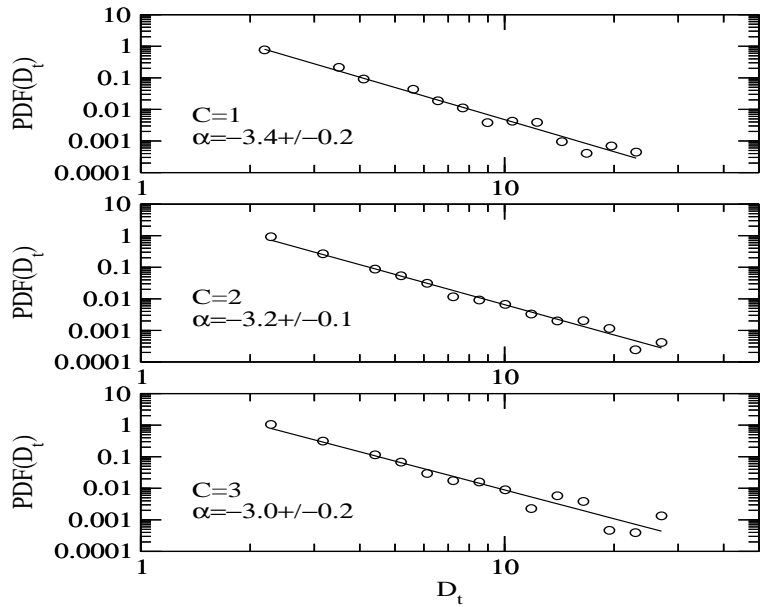}}
\caption{ Distribution of the duration of the Dow Jones index.}
\label{fig12}
\end{figure}

\begin{figure}
\vspace{1cm}
\centerline{\epsfig{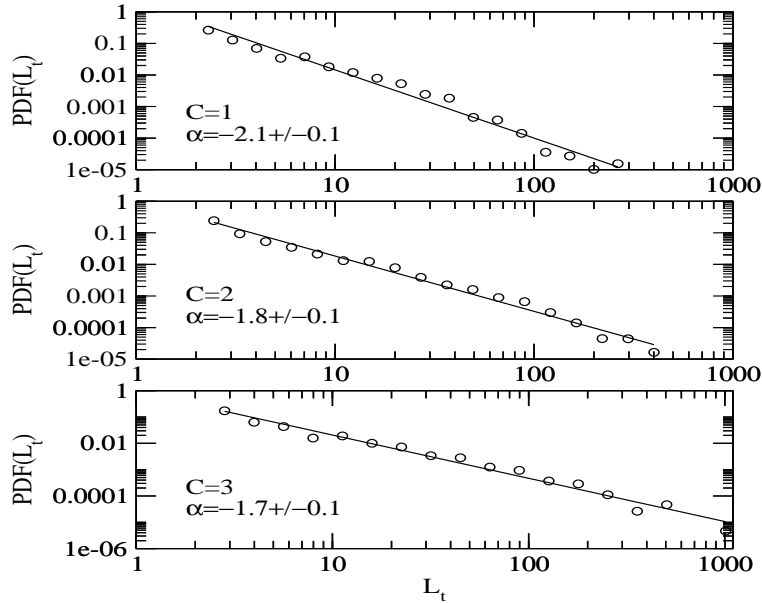}}
\caption{ Distribution of laminar times for the Dow Jones index.}
\label{fig13}
\end{figure}

We also investigate the possibility of differences between high frequency
 data and daily closures by considering
a sample of $2^{14}$ daily closures 
of the Dow Jones index, from 2/2/1939 to 13/4/2004.
The power law behaviour is consistent with that found for 
the high frequency data, as shown
in Figs.~\ref{fig11}, ~\ref{fig12} and ~\ref{fig13}. 

\section{Discussion}

Similar power law behaviour for $V$, $D_{t}$ and $L_{t}$ has been 
found in the context of solar flaring~\cite{Boffetta99} and  
in geophysical time-series~\cite{Kovacs01}.
In the case of solar flaring, Boffetta et. al~\cite{Boffetta99} have
shown that the characteristic distributions found 
empirically are more similar to the dissipative behaviour of
the shell model for turbulence~\cite{Bohr,Giuliani98} than to SOC.
On the other hand the intermittency in turbulent flows discussed in
 Sec~\ref{sec:wavelet} is believed to be the result of a non-linear
 energy cascade that generates non-Gaussian events at small
 scales~\cite{Frisch} where the shape of the PDF is extremely
 leptokurtic. At larger   scales the spatial correlation decreases and
 the PDF converges toward a  Gaussian. These features imply the
 absence of global self-similarity -- which, as we have noted, is  
an intrinsic component of
 SOC models~\cite{Carbone02}.

Freeman {\em et. al}~\cite{Freeman00} have argued that, although an 
 exponential distribution holds for classical sandpile
models, there exist some non-conservative modifications of the
BTW models in which departures from an exponential behaviour 
for the $L_{t}$ distribution~\cite{Christensen92,Olami92,Hwa92,Chau92}
are observed in the presence of scale-free dynamics for other relevant parameters.
The question remains whether or not these systems are
still in a SOC state~\cite{Freeman00}. 
If we assume that the power law scaling of the laminar times corresponds to a
breakdown of self-organized criticality, then we
face the problem of how to explain the observed scale-free behaviour
of the non-conservative models.
This ambiguity can be resolved if
we assume that the system is in a {\em near-SOC} state, that is the
scaling properties of the system are kept even if it is not
exactly  critical and temporal correlations
may be present~\cite{Freeman00,Carvalho00}.  
Another possible scenario is related with the existence of temporal
correlations in the driver~\cite{DeLosRios97,Sanchez02}.
In this case the power law behavior of the 
waiting time distribution would be explained and the realization of a SOC state
preserved~\cite{DeLosRios97,Sanchez02}.

\section{Conclusions}
In the present work we have investigated empirically the possible 
relations between the theory of self-organized criticality and the 
stock market. The existence of a SOC state for the market would
be of great theoretical importance, as this would impose some constraints
on the dynamics of this complex system. A bounded attractor in the state
space would be implied. Moreover, we would have a better understanding of 
business cycles. 

From the wavelet analysis on a sample of high frequency 
data for the Nasdaq100 index, we have found that the
behaviour of high activity periods, or avalanches, 
is characterized by power laws in   
the size, duration and laminar times. The power laws in the avalanche size
and duration are a characteristic feature of a critical underlying dynamics in
the system, but this is not enough to claim the self-organized critical state.
 In fact the power law behavior in the laminar time distribution implies a
 memory process in the triggering driver that is absent in the classical BTW models, 
where an exponential behavior is expected. This does not rule out completely
 the hypothesis of underlying self-organized critical dynamics in the market. 
Non-conservative systems, as for the case of the stock market, near the SOC state can
still show  power law dynamics even in presence of temporal correlations of
the avalanches~\cite{Freeman00,Carvalho00}. Another possible explanation is that the 
memory process, possibly chaotic,
 is intrinsic in the driver. In this case the power law behavior of the 
waiting time distribution would be explained and the realization of a SOC state
preserved~\cite{DeLosRios97,Sanchez02}. 

These findings extend beyond the Nasdaq100 index analysis.
A similar  quantitative behaviour has been
 observed in the S\&P ASX50 high frequency data for the Australian market 
and the daily closures of the Dow Jones index for the American market.
 At this point it is important to stress that this
 does not imply that all the markets must display the same identical 
characteristics. In the case of a near-SOC dynamics, for example, the power law
shape of the distribution can be influenced by the degree 
of dissipation of the system which may change from market 
to market, implying a non-universal behaviour.

In conclusion, a definitive relation between SOC theory and the stock market has
not been found. Rather, we have shown that a  memory process is related 
with periods of high activity. The memory could result from some 
kind of dissipation of
information, similar to turbulence, or possibly a chaotic driver applied to the
self-organized critical system. Of course, a combination of the two processes can
also be possible. Our future work will be devoted 
to the study of new tests for self-organized criticality and the implementation
of numerical models~\cite{Bartolozzi04}.

\section*{Acknowledgements}
This work was supported by the Australian Research Council.

\end{document}